# Towards rational design of carbon nitride photocatalysts: Identification of cyanamide "defects" as catalytically relevant sites

Vincent Wing-hei Lau, Igor Moudrakovski, Tiago Botari, Simon Weinburger, Maria B. Mesch, Viola Duppel, Jürgen Senker, Volker Blum, Bettina V. Lotsch

## Abstract

The heptazine-based polymer melon (also known as graphitic carbon nitride, g-C3N4), is a promising photocatalyst for hydrogen evolution. Nonetheless, attempts to improve its inherently low activity are rarely based on rational approaches due to a lack of fundamental understanding of its mechanistic operation. Here, we employ molecular heptazine-based model catalysts to identify the cyanamide moiety as a photocatalytically relevant "defect". We exploit this knowledge for the rational design of a carbon nitride polymer populated with cyanamide groups, yielding a material with 12- and 16-times the hydrogen evolution rate and apparent quantum efficiency (400 nm), respectively, compared to the benchmark melon. Computational modelling and material characterization suggest this moiety improves co-ordination (and, in turn, charge transfer kinetics) to the platinum co-catalyst and enhances the separation of the photo-generated charge carriers. The demonstrated knowledge transfer for rational catalyst design presented here provides the conceptual framework for engineering high performance heptazine-based photocatalysts.

## Introduction

Photocatalytic hydrogen evolution is considered to be a promising technology for the direct capture and storage of solar energy, since the energy stored in the hydrogen-hydrogen bond can be extracted in high efficiency without producing environmentally harmful by-products.[1] Numerous materials[2] have been reported since Honda's and Fujishima's seminal report[3] on water splitting photocatalyzed by titanium dioxide, yet none of them are at a stage for industrial deployment. The common disadvantages with most of these photocatalytic materials are: 1) low activity under visible light; 2) costliness associated with rare elements and/or synthetic processes, and 3) lack of chemical stability.

One promising photocatalytic material is melon,[4-6] also referred to erroneously in the literature as graphitic carbon nitride (g-$C_3N_4$).[7] It is easily prepared from inexpensive precursors (urea, dicyandiamide or melamine), has a suitable band gap and potential for water reduction/oxidation under visible light,[5,8,9] good chemical stability and, finally, molecular tunability. Its low activity is addressed primarily through texturization for increased surface area,[10-12] copolymerizing with organic[13-15] or inorganic[16-18] dopants for tuning its optoelectronic properties, compositing with (semi)conductors for enhanced charge separation,[6,19,20] and employing hetero- or homogeneous co-catalysts to improve interfacial charge transfer.[21-25] Our approach is to gain an intimate understanding of the structure-activity relationship of these carbon nitride materials, then transferring this knowledge for the rational design of improved, i.e. inherently more active, photocatalysts. Following the accepted postulate that "…surface terminations and defects seem to be the real active sites…",[6] we have previously demonstrated that oligomers outperform polymers in terms of photocatalytic hydrogen evolution rate.[26] Here, we aim to elucidate the identities of the catalytically relevant surface terminations of carbon nitrides by employing heptazine-based molecules as model catalysts. This methodology, employed to circumvent complication from surface heterogeneity and structural ambiguity in heterogeneous catalysis research,[27-31] is particularly suitable for studying melon which, due to its amorphous nature and lack of solubility, could only be structurally resolved nearly two centuries after its discovery. Its photophysical properties resemble that of a quasi-monomer, suggesting exciton localization within each heptazine unit of the polymer,[32,33] thereby justifying the use of heptazine molecules as the model candidates. In this work, we first screened for photocatalytic activity for hydrogen evolution in the heptazine molecules and complexes shown in Scheme 1, which differ in the functional group at the 2, 5, and 8 positions of the heptazine core. Having identified catalytically relevant moieties, we further demonstrate the process of rational catalyst design by introducing one of these functionalities – the cyanamide group – into the polymer chain to enhance its intrinsic photocatalytic activity.

## Results & Discussions

**Model Photocatalyst**

Functional groups that are identified as potentially relevant for catalysis include not only amines (primary, secondary or tertiary), but also the cyanamide moiety and oxygen-bearing groups (Scheme S1). Cyanamide may occur due to incomplete heptazine cyclization based on some proposed mechanisms for heptazine formation or thermal depolymerization,[34-36] while oxygen-bearing groups may be present from impurities in the precursors (e.g. ureidomelamine is a significant impurity in melamine[37]) or oxygen contamination through trace water during high temperature syntheses in air. We emphasize that these groups have largely been overlooked to date as catalytically relevant in the literature, and most publications generally depict *g*-$C_3N_4$ as an idealized 2D polymer of heptazine units bridged (2° and 3°) and terminated solely by amines (1°). Note that the photocatalytic models were screened only for the presence of activity but were not compared for their hydrogen evolution rates quantitatively, since the models differ in their spectral profile and solubility/dispersibility in aqueous solution.

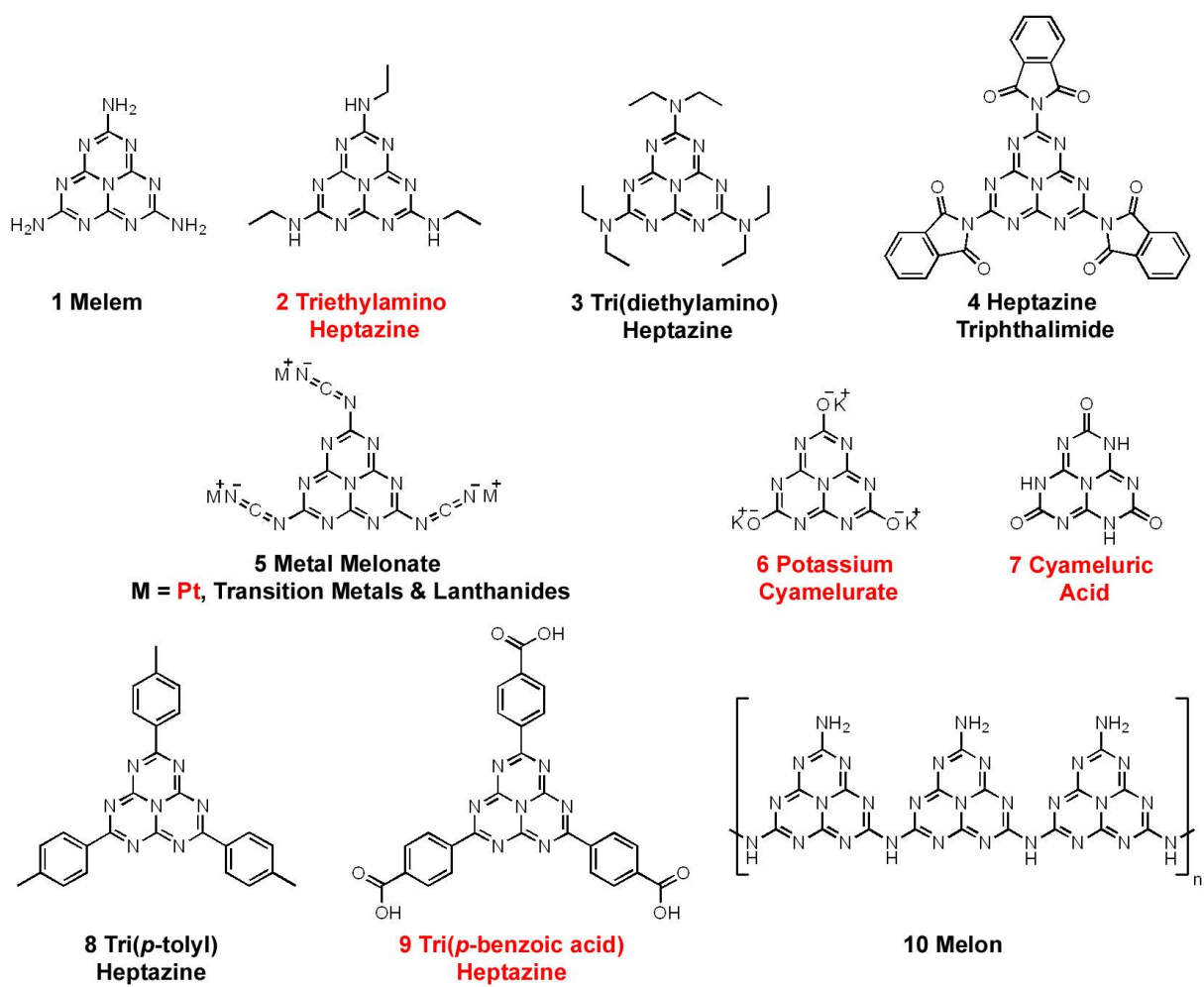

Scheme 1 Molecular structures of the molecular models used in this study. The ones labelled red exhibited photocatalytic activity for hydrogen evolution under standard conditions. Sample codes for the melonate complexes with different metals are summarized in the SI. Both melem and melon are denoted as inactive based on their crystalline counterparts, rather than the structurally ambiguous amorphous versions (see supporting discussion regarding structure on page S11).[26]

Using the standard protocol as outlined in the SI, photocatalytic activity was observed (Figure S5) for the secondary amine (**2**), the cyanamide-platinum complex (**5k**, with and without added $H_2PtCl_6$), and the oxygen-containing models (**6** and **7**), thus justifying our hypothesis that functionalities from incomplete heptazine cyclization and impurities can also be catalytically relevant and may, in fact, be some of the catalytically active "defects" cited in the literature.[6] Activity was also observed for the heptazine tribenzoic acid (**9**) but not for the tolyl equivalent (**8**), suggesting the necessity of a coordinating group. While the above catalytically active species represent the first molecular model compounds for carbon nitride photocatalysts, Pt-melonate is distinct in its well-defined coordination mode and the fact that it produces hydrogen both with and without addition of $H_2PtCl_6$. Pt-melonate therefore provides a unique platform for investigating charge transfer from the heptazine core to the solution, via the platinum centre, for hydrogen evolution.

From the screening of these models, we attempted to identify some predictive descriptors based on properties intrinsic to the molecule itself, namely frontier orbitals and their energy levels, which may lead to the presence of activity, using density functional theory (DFT) adapted[38] to approximate higher-level $G_0W_0$ quasiparticle calculations for the energy levels of the isolated models (FHI-aims all-electron code;[39,40] see Figure S13, page S7 and S20 for computational details). In our calculations, the photocatalytic activity was not found to correlate with the HOMO/LUMO locations, the energy levels, nor the energy gaps. The absence of a correlation is most evident for the melem-derived series of models **1**, **2** and **3**, which differ by the successive attachment of ethyl groups to the amine side group but are electronically very similar, suggesting that other intrinsic or extrinsic factors such as intermolecular interactions and steric hindrances are at play. Combining all the screening results, we infer that a required (though not necessarily sufficient) criterion for catalytic activity may be the presence of a coordinating group, which provides the ligating linkage to the platinum centers to facilitate the transfer of the photoexcited charge from the heptazine core. This criterion however can be complicated by intra-/intermolecular bonding in the condensed phase, where hydrogen bonding or steric hindrance locks up the availability of the ligating moieties. The models where this may be the case are crystalline melem and melon, where all amine groups are blocked by hydrogen bonding, as well as tri(diethylamino)heptazine, where the ethyl groups hinder coordination between the 3° amine and the platinum.

**Cyanamide Functionalization of Melon**

Armed with these insights, we next demonstrate the rational design of a highly active carbon nitride photocatalyst by deliberately populating the polymer with one such catalytically relevant group identified in the screening experiment, namely the cyanamide moiety. The rationale for choosing this particular functional group comes from MALDI-TOF measurements, which suggest that cyanamide groups are indeed present already in native amorphous melem synthesized from melamine under argon at ambient pressure (Fig. S3-4 and Table S1). The NCN-moiety may therefore be innate in small amounts to carbon nitrides when synthesized under certain conditions and result from polymerization/depolymerization equilibria involving cyanamide as suggested for the formation of melem (Scheme S1).

In our prototype design, functionalization of melon with cyanamide was achieved following the KSCN salt melt synthesis for potassium melonate, except that the yellow insoluble solid was isolated, following the mechanism proposed in Scheme S2. Sulfur, though detected by ICP-AES in very small amounts (0.17 wt%, see Table S2) but not by XPS, did not seem to adversely affect the photocatalytic activity (see below). This residue (notated henceforth as "KSCN-treated melon") exhibited a steady hydrogen evolution rate of 24.7 µmol h$^{-1}$ after more than 70 h (31.2 µmol h$^{-1}$ in the first 4 h) of AM1.5 irradiation at an optimum platinum loading of 8 wt% using $H_2PtCl_6$ as the platinum source and 10 vol% of aqueous methanol as electron donor (Figure 1; platinum optimization in Figure S14). Apparent quantum efficiencies (AQE) were estimated to be 9.3% under 400 nm irradiation and 0.34% across the entire AM 1.5 spectrum. In accordance with our hypothesis, this vastly outperformed amorphous melon, which has a hydrogen evolution rate of 2.0 µmol h$^{-1}$ at 1 wt% optimized platinum loading under identical conditions and an AQE of 0.5% at 400 nm and 0.027% for AM 1.5. Negligible activities under 500 nm irradiation for both samples are consistent with their absorption profile (Figure 1). While an internal quantum efficiency of 26.5% have been reported for heptazine[41] photocatalysts recently, these experiments employed triethanolamine, which as an electron donor increases the hydrogen evolution rate by 3.5 times compared to methanol (Figure S15) on account of its more negative redox potential.[42] We nonetheless did not select triethanolamine (nor the even more active sodium oxalate, see Figure 1d) to optimize our AQE, since it suffers from some disadvantages (e.g. light sensitivity, optical impurity, E° = -600 mV) not present for methanol (E° = +200 mV). In stating these, we consider our AQE of 9% competitive with the above literature values, since they do not exceed our value by more than 3 times. In this sense, the activities observed for KSCN-treated melon are among the best reported for carbon nitride photocatalysts, although in our case neither compositing nor texturization was applied. Negligible change of their optoelectronic properties based on spectral profile and Fermi levels (valence band region of XPS in Figure S16) rule the changes in band energies out as the principal factor for KSCN-treated melon's high performance.

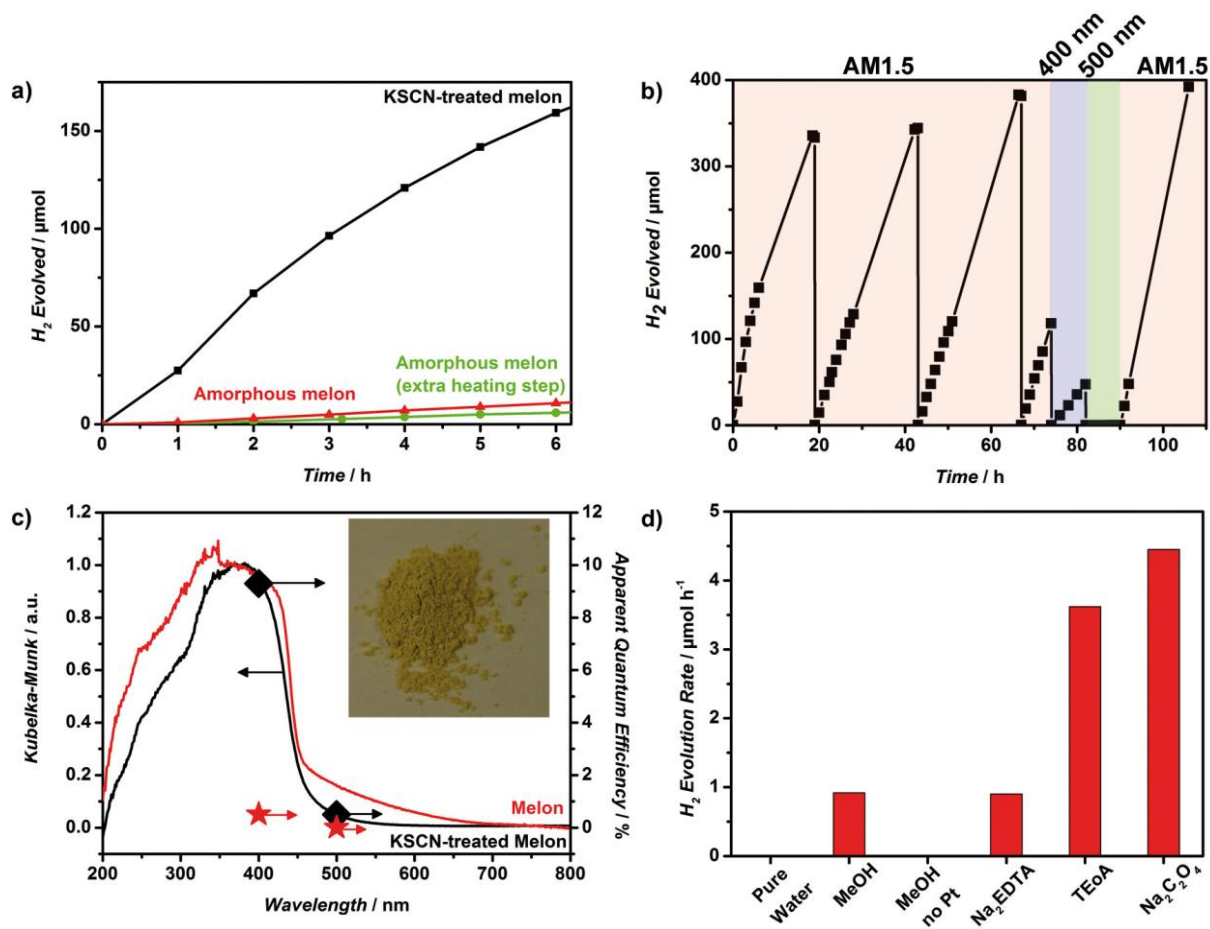

Figure 1 a) Photocatalytic hydrogen evolution of KSCN-treated melon compared with amorphous melon, and amorphous melon that has undergone an extra heating step identical to the KSCN-treated melon (400 °C for 1 h, then 500 °C for 30 min) for the first six hours and for b) 100+ h. After every overnight run, the headspace of the reactor was evacuated and flushed with argon. Methanol (500 µL) was added on the 43$^{rd}$ and 67$^{th}$ hour, and on the 74$^{th}$, 82$^{nd}$, and the 90$^{th}$ hour (200 µL). c) Action spectra of KSCN-treated melon and melon showing photocatalytic activity over two wavelengths; inset shows a photograph of the KSCN-treated melon. d) Comparison of hydrogen evolution rate in different electron donors, using ethylenediamine tetraacetic acid (disodium salt, Na$_2$EDTA), methanol (MeOH), triethanolamine (TEoA), and sodium oxalate at the optimized platinum loading. Reaction conditions: catalyst (20 mg) dispersed in aqueous solution of the donor (20 mL, 50 mM) and H$_2$PtCl$_6$ (40 µL of 8 wt% aqueous solution) under AM 1.5 irradiation. Control experiments conducted without electron donor ("pure water") and with methanol (10 vol%), but without addition of platinum co-catalyst ("MeOH no Pt") were also performed.

To explore the structural and compositional features of KSCN-treated melon leading to its high photocatalytic performance, we characterized the material by MAS ssNMR, FTIR and Raman spectroscopies, XRD, SEM and TEM (Figure 2), as well as elemental analysis (Table S2), TGA-MS (Figure S17), XPS (Figure 3) and physisorption (Figure S18). Note that NMR spectroscopy was performed on the material prepared from both KSCN and enriched KS$^{13}$C$^{15}$N (FTIR comparing the two, see Figure S19). The heptazine core is confirmed by its characteristic IR vibration at 804 cm$^{-1}$,[43-45] the Raman ring vibrations in the range 1653–1160 cm$^{-1}$,[46] the $^{13}$C NMR shifts at 157 (C2) and 163 ppm (C3), and the $^{15}$N shifts for the central (N3) and peripheral nitrogens (N4) at -225 and -160 – -205 ppm, respectively, as well as the sp$^2$ carbon (288.4 eV) and nitrogen (398.8 eV) signals in the XPS. Its polymeric nature is shown by the presence of the bridging 2° amine as observed in the C–N IR bending mode (1311 and 1221 cm$^{-1}$),[47-49] the $^{15}$NH (N2) NMR signal at -243 ppm, and the XPS signal at 401.1 eV. The central N3 $^{15}$N NMR signal of the heptazine ring at -225 ppm, identical to the polymer[7] and dissimilar to the monomer[43] (-235 ppm) and oligomers (-225<δ<-235 ppm)[26] as well as the absence of blue-shift in the UV-Vis spectrum are evidence for the polymeric nature of KSCN-treated melon.[26] The cyanamide group is evidenced by the IR and Raman band at 2177 cm$^{-1}$ assigned to vibrations involving C≡N stretch,[50] and the corresponding $^{13}$C signal is observed at 118.2 ppm (C1) and $^{15}$N signals at -175 (N1) and -276 ppm (N5), consistent with literature values for melonate (potassium[46] and other[51] metal salts), organic cyanamide compounds,[52] and tricyanomelaminates.[53,54] The reasons for the lower than expected intensities for the cyanamide NMR signals are discussed in greater depth in the supporting information, but the postulated structure is nonetheless supported by elemental analyses, with the C:N atomic ratio found to be 0.70 and roughly 40% potassium exchanged with protons (Table S2).

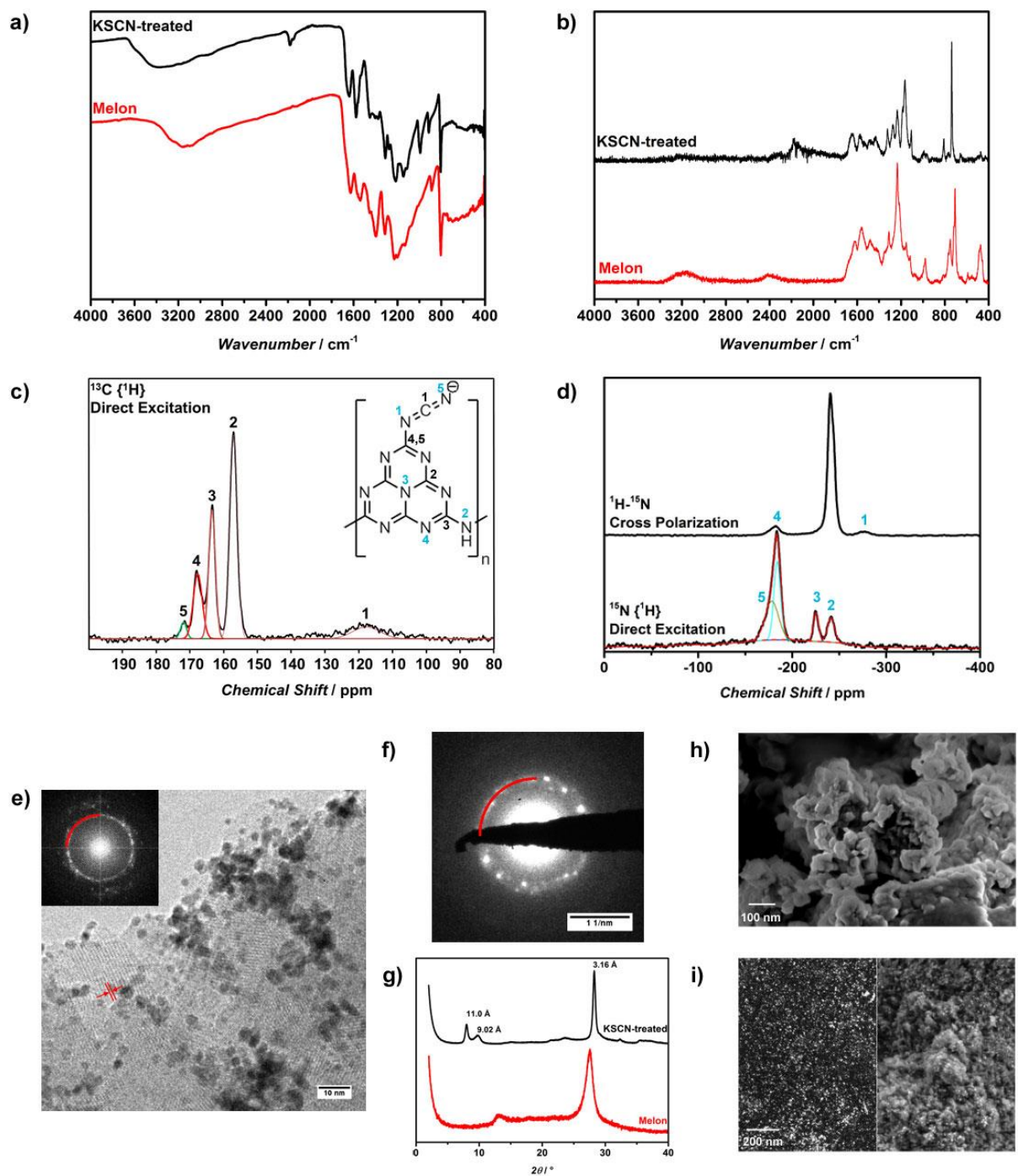

Figure 2 Characterizations of KSCN-treated melon compared with amorphous melon: a) FTIR; b) Raman; solid-state MAS c) $^{13}$C and d) $^{15}$N spectra of KSCN-treated melon. For the $^{15}$N spectrum, melon was treated in isotope enriched KS$^{13}$C$^{15}$N. A summary of the deconvolution and integration of the $^{13}$C spectrum is given in Table S3. Inset shows the proposed structure and NMR assignments (carbon 4 and 5 refer to H$^+$ and K$^+$ as counter-ion, respectively); e) bright field TEM image of the spent catalyst with the in-situ deposited platinum particles of diameter 2–4 nm and the FFT of the image (inset); f) SAED pattern of the spent catalyst; all markings in red for both e) and f) indicate a lattice spacing of 11 Å; g) XRD pattern; h) SEM before catalysis; i) SEM of the spent catalyst imaged with ESB (energy and angle selective) backscattered electron (left) and imaged with secondary electron (right) detector; the ESB backscattered image shows the photo-reduced platinum as bright spots.

Compared to the XRD of amorphous melon with the stacking and in-plane periodicities[7,55] at 27.56° 2θ (3.23 Å) and 12.8° 2θ (6.9 Å), respectively, KSCN-treated melon has a denser layering (28.26° 2θ, 3.16 Å) and exhibits in-plane periodicities at 11.0 Å (8.03° 2θ) and 9.02 Å (9.82° 2θ). The 11 Å lattice fringes are also observed in the bright field images and its FFT, as well as the selected area electron diffraction (SAED) analysis. Sorption measurements indicate most of the pore volume is in the micro- and low mesoporosity range (Figure S17 for pore size distribution histogram) with the larger mesopores attributed to inter-particle spacing. Although the BET surface area and pore volume are respectively 3.4 and 2.7 times higher than amorphous melon's (Table S4), the intrinsic activity of KSCN-treated melon based on normalization of activity to BET surface area still outperforms amorphous melon's activity – 22.2 μmol $H_2$ $h^{-1}$ $m^{-2}$ versus 7.5 μmol $H_2$ $h^{-1}$ $m^{-2}$ – which we attribute to the rational insertion of the cyanamide group.

**Catalytic Role and Evolution of the Cyanamide Moiety**

Characterizations of the spent catalyst after 100+ h of irradiation (Figure 3, NMR in Figure S21 and supporting discussion on page S30) indicate partial hydrolysis of the cyanamide moiety to urea and may account for the drop of the hydrogen evolution rate of 31.2 μmol $h^{-1}$ in the first 4 h to the stable rate of 24.7 μmol $h^{-1}$ thereafter. The FTIR, NMR and XPS spectra of the spent catalyst are nearly identical to those of melon terminated by urea which can be prepared by acid hydrolysis of the KSCN-treated melon (see SI). Our rough estimate, based on changes in the ratio of FTIR intensities of the cyanamide (2181 $cm^{-1}$) to the heptazine signal (805 $cm^{-1}$), is that 67% of the cyanamide has been hydrolyzed in the spent catalyst after 100+ h. We nonetheless emphasize that for this prototype catalyst designed from our model screening, the final steady rate for hydrogen evolution is still 80% that of the initial rate and significantly higher than that of amorphous melon, as outlined above. We rationalize this observation by the continued strong interaction between Pt and the carbon nitride at the Pt-$CN_xO_y$ heterointerface, even after hydrolysis. This is supported by the presence of both $Pt^0$ and $Pt^{II}$ in the XPS of the spent KSCN-treated melon, as well as that of melon, which is consistent with literature observations.[56] However, the lower than expected binding energies (B.E.) for $Pt^0$ at 71.1-71.2 eV and for $Pt^{II}$ at 72.4 eV are more akin to those of platinum on strongly interacting supports like $TiO_2$ or $CeO_2$ (70.4-70.8 eV for $Pt^0$ and 71.5-72.5 eV for $Pt^{II}$)[57-62] rather than on less interacting supports like $SiO_2$ or carbon (71.6-71.9 eV for $Pt^0$ and >72.7 eV for $Pt^{II}$).[59,60,63-65] Together with the minute shifts to higher B.E. for the $N_{1s}$ and $C_{1s}$ signals in the spent melon, these XPS results allude to a metal-support interaction (MSI) in these carbon nitride photocatalysts with polarization of electron density from the support to the platinum, based on the direction of these shifts. Shifts of all signals larger than spectrometer resolution, except for the adventitious carbon signal used for calibration, lend weight that the observed changes are not calibration errors. This MSI effect has also been attributed to improved platinum electrocatalytic activities[60,62,66,67] for reactions such as the oxygen reduction reaction. Although the carbon nitrides are not as interacting as $TiO_2$ and $CeO_2$ based on the magnitudes of the shifts, this MSI effect provides a consistent rationale to the necessity of a coordinating group in the heptazine models or of "defects" in amorphous melon to exhibit photocatalytic activity. Importantly, the role of such ligating groups or defects may thus be to facilitate interactions between the polymer and the co-catalyst for efficient charge transfer.

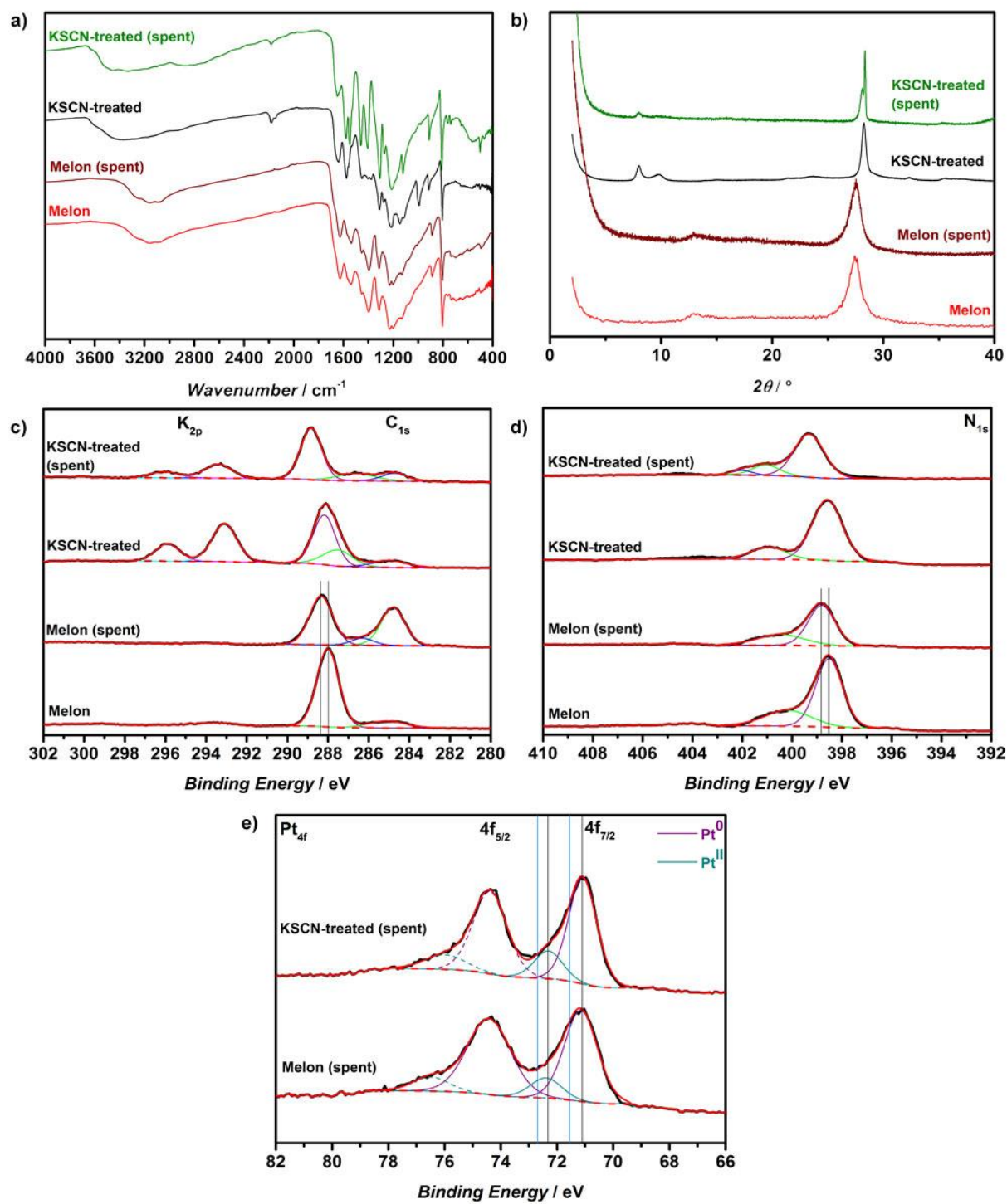

Figure 3 Characterization of the KSCN-treated melon before and after the 100+ h photocatalytic experiment from Figure 1, compared with amorphous melon: a) FTIR; b) XRD, and XPS in the regions of c) $K_{2p}$ and $C_{1s}$, d) $N_{1s}$, and e) $Pt_{4f}$. Markings in the $C_{1s}$ and $N_{1s}$ XPS spectra indicate the slight shifts observed between the pristine and spent photocatalysts. In the $Pt_{4f}$ spectrum, for comparison of the B.E. of the platinum on the melon samples, light blue lines are drawn to show the B.E. (adjusted for the different calibrations used) of $Pt^0$ on graphite[67] and $Pt^{2+}$ on $SiO_2$,[70] both non-strongly interacting supports.

**Computational Modelling of the Cyanamide Functional Group**

In addition to its promotional effect, we explored computationally whether this anionic moiety increases local fluctuations of the electrostatic potential within the polymer, thus enhancing the charge separation. As an indicator of the possible charge separation tendency, we calculated the HOMO/LUMO locations for three different terminations of the same heptazine pentamer conformation using DFT as above (page S30 for computational details, Figures S22–S24, energy levels summarised in Table S5). Figure 4 shows the resulting orbitals for the pentamer terminated with $NH_2$, proton-terminated NCN, or potassium-coordinated NCN side groups, i.e., three overall neutral molecules with increasing ionicity of the side group and counter charge. Obviously, the HOMO and LUMO for the case with $K^+$ as the cation are well separated, significantly more localized, and contained within different heptazine units of the pentamer, than for the hydrogen-terminated case or the termination by $NH_2$. We attribute the differences in preferential electron or hole locations on the pentamer to the stronger local electrostatic potential differences on different heptazine units due to the locations of the $K^+$ cations. One consequence is the spatial separation of the electron-hole pair, thereby reducing their recombination and increasing their probability of interfacial charge transfer for the desired redox reaction. These results are similar to semiconductors with a built-in electric field, which lead to better charge separation and hence higher activity.[68,69] More significantly, the result implies that introducing an ionic moiety may be a valid strategy for enhancing charge separation in carbon nitrides. As exciton localization has been considered to be the major limitation to the photocatalytic efficiency for carbon nitride-type materials,[70] our current work may open new avenues for rationally improving the photocatalytic activity of such systems.

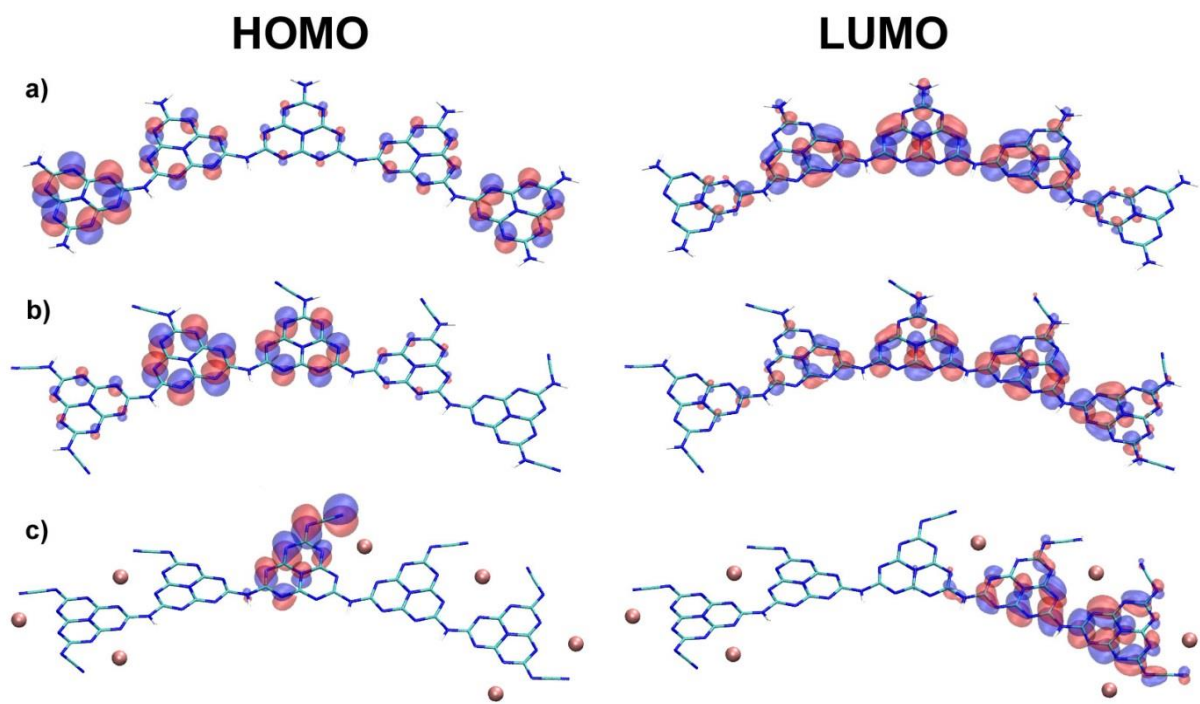

Figure 4 HOMO and LUMO distribution of: a) melon pentamer and melon pentamer with all –$NH_2$ replaced by $NCN^-$ charge balanced by b) proton and c) potassium, calculated using the ic-PBEh functional (see page S7 and S31 for computational details).

## Conclusion

Using heptazine-based molecular models for screening photocatalytic activity, we have identified catalytically relevant functional groups in amorphous melon. We then exploited this knowledge for rational catalyst design by populating amorphous melon with one of these groups, the anionic cyanamide moiety. This prototype catalyst evolves hydrogen at a stable rate at least 12 times higher than the benchmark melon and 18 times in terms of apparent quantum efficiency at 400 nm. Characterization of the catalyst after photocatalysis and computational experiments suggest two roles of this moiety: 1) enhancing co-catalyst interactions and thus facilitating interfacial electron transfer to the platinum centers, and 2) improving the separation of the photoexcited charges through built-in electrostatic potential differences across the heptazine polymer. In demonstrating this case of "defect engineering" with cyanamide, we emphasize that such moieties may be present in the prototype polymer melon formed under certain synthesis conditions, based on the TGA-MS and MALDI-TOF analyses of amorphous melem and melon. These findings provide the rationale for the wide variation of photocatalytic activity as the synthesis conditions and the precursors used are changed, namely the differences in the amount and type of catalytically relevant "defects" formed in the resulting material. Investigating the promotional role of such "defects" thus offers a yet unexplored avenue to improving the *intrinsic* photocatalytic activity of polymeric carbon nitride. The methodologies, mechanistic insights, and process of rational catalyst design demonstrated herein can be adapted to studying other light-driven reactions (oxygen evolution or $CO_2$ reduction), as well as exploring other possible functional groups non-native to melon, which ultimately would provide the blueprint for the design of highly active, customizable heptazine-based photocatalysts.

## Acknowledgement

V.W.-h.L. gratefully acknowledges a postdoctoral scholarship from the Max Planck Society. T.B. acknowledges support by grant #2014/11986-7, São Paulo Research Foundation (FAPESP). This work was supported by the Deutsche Forschungsgemeinschaft (projects LO1801/1-1, SE1417/5-1), the Max Planck Society, the cluster of excellence Nanosystems Initiative Munich (NIM), the Fonds der Chemischen Industrie (FCI), and the Center for Nanoscience (CeNS). The authors would like to thank Dr. Stephan Rauschenbach for assistance in the preparation and analysis of the samples for MALDI-TOF measurements. The authors would also like to thank Ms. Marie-Luise Schreiber for the elemental analyses, Mr Quirin Axthammer for the Raman measurements, Dr. Mitsuharu Konuma for the XPS analyses, and Mr. Olaf Alberto von Mankowski for the argon sorption measurements.

# References

(1) Rand, D. A. J.; Dell, R. M. *Hydrogen Energy: Challenges and Prospects*; RSC Publishing: Cambridge, 2008.
(2) Xing, J.; Fang, W. Q.; Zhao, H. J.; Yang, H. G. *Chem. Asian J.* **2012**, *7*, 642
(3) Fujishima, A.; Honda, K. *Nature* **1972**, *238*, 37
(4) Maeda, K.; Wang, X.; Nishihara, Y.; Lu, D.; Antonietti, M.; Domen, K. *J. Phys. Chem. C* **2009**, *113*, 4940
(5) Wang, X.; Maeda, K.; Thomas, A.; Takanabe, K.; Xin, G.; Carlsson, J. M.; Domen, K.; Antonietti, M. *Nat. Mater.* **2009**, *8*, 76
(6) Wang, Y.; Wang, X.; Antonietti, M. *Angew. Chem. Int. Ed.* **2012**, *51*, 68
(7) Lotsch, B. V.; Döblinger, M.; Sehnert, J.; Seyfarth, L.; Senker, J.; Oeckler, O.; Schnick, W. *Chem. Eur. J.* **2007**, *13*, 4969
(8) Xu, Y.; Gao, S.-P. *Intl. J. Hydrogen Energy* **2012**, *37*, 11072
(9) Butchosa, C.; Guiglion, P.; Zwijnenburg, M. A. *J. Phys. Chem. C* **2014**, *118*, 24833
(10) Chen, X.; Jun, Y.-S.; Takanabe, K.; Maeda, K.; Domen, K.; Fu, X.; Antonietti, M.; Wang, X. *Chem. Mater.* **2009**, *21*, 4093
(11) Niu, P.; Zhang, L.; Liu, G.; Cheng, H.-M. *Adv. Func. Mater.* **2012**, *22*, 2763
(12) Sun, J.; Zhang, J.; Zhang, M.; Antonietti, M.; Fu, X.; Wang, X. *Nat. Commun.* **2012**, *3*, 1139
(13) Zhang, J.; Chen, X.; Takanabe, K.; Maeda, K.; Domen, K.; Epping, J. D.; Fu, X.; Antonietti, M.; Wang, X. *Angew. Chem. Int. Ed.* **2010**, *49*, 441
(14) Zhang, J.; Zhang, G.; Chen, X.; Lin, S.; Möhlmann, L.; Dołęga, G.; Lipner, G.; Antonietti, M.; Blechert, S.; Wang, X. *Angew. Chem. Int. Ed.* **2012**, *51*, 3183
(15) Schwinghammer, K.; Tuffy, B.; Mesch, M. B.; Wirnhier, E.; Martineau, C.; Taulelle, F.; Schnick, W.; Senker, J.; Lotsch, B. V. *Angew. Chem. Int. Ed.* **2013**, *52*, 2435
(16) Liu, G.; Niu, P.; Sun, C.; Smith, S. C.; Chen, Z.; Lu, G. Q. M.; Cheng, H.-M. *J. Am. Chem. Soc.* **2010**, *132*, 11642
(17) Wang, Y.; Di, Y.; Antonietti, M.; Li, H.; Chen, X.; Wang, X. *Chem. Mater.* **2010**, *22*, 5119
(18) Lin, Z.; Wang, X. *Angew. Chem. Int. Ed.* **2013**, *52*, 1735
(19) Zheng, Y.; Jiao, Y.; Zhu, Y.; Li, L. H.; Han, Y.; Chen, Y.; Du, A.; Jaroniec, M.; Qiao, S. Z. *Nat. Commun.* **2014**, *5*, 3783
(20) Liu, J.; Liu, Y.; Liu, N.; Han, Y.; Zhang, X.; Huang, H.; Lifshitz, Y.; Lee, S.-T.; Zhong, J.; Kang, Z. *Science* **2015**, *347*, 970
(21) Dong, J.; Wang, M.; Li, X.; Chen, L.; He, Y.; Sun, L. *ChemSusChem* **2012**, *5*, 2133
(22) Cao, S.-W.; Liu, X.-F.; Yuan, Y.-P.; Zhang, Z.-Y.; Fang, J.; Loo, S. C. J.; Barber, J.; Sum, T. C.; Xue, C. *Phys. Chem. Chem. Phys.* **2013**, *15*, 18363
(23) Hong, J.; Wang, Y.; Wang, Y.; Zhang, W.; Xu, R. *ChemSusChem* **2013**, *6*, 2263
(24) Cao, S.-W.; Yuan, Y.-P.; Barber, J.; Loo, S. C. J.; Xue, C. *Appl. Surf. Sci.* **2014**, *319*, 344
(25) Caputo, C. A.; Gross, M. A.; Lau, V. W.; Cavazza, C.; Lotsch, B. V.; Reisner, E. *Angew. Chem. Int. Ed.* **2014**, *53*, 11538
(26) Lau, V. W.-h.; Mesch, M. B.; Duppel, V.; Blum, V.; Senker, J.; Lotsch, B. V. *J. Am. Chem. Soc.* **2015**, *137*, 1064
(27) Muetterties, E. L. *Science* **1977**, *196*, 839
(28) Moskovits, M. *Acc. Chem. Res.* **1979**, *12*, 229
(29) Marks, T. J. *Acc. Chem. Res.* **1992**, *25*, 57
(30) Quadrelli, E. A.; Basset, J.-M. *Coord. Chem. Rev.* **2010**, *254*, 707
(31) Copéret, C.; Chabanas, M.; Saint-Arroman, R. P.; Basset, J.-M. *Angew. Chem. Int. Ed.* **2003**, *42*, 156

(32) Tyborski, T.; Merschjann, C.; Orthmann, S.; Yang, F.; Lux-Steiner, M.-C.; Schedel-Niedrig, T. *J. Phys. Condens. Matter* **2012**, *24*, 162201

(33) Merschjann, C.; Tyborski, T.; Orthmann, S.; Yang, F.; Schwarzburg, K.; Lublow, M.; Lux-Steiner, M.-C.; Schedel-Niedrig, T. *Phys. Rev. B* **2013**, *87*, 205204

(34) Hosmane, R. S.; Rossman, M. A.; Leonard, N. J. *J. Am. Chem. Soc.* **1982**, *104*, 5497

(35) Shahbaz, M.; Urano, S.; LeBreton, P. R.; Rossman, M. A.; Hosmane, R. S.; Leonard, N. J. *J. Am. Chem. Soc.* **1984**, *106*, 2805

(36) Schwarzer, A.; Saplinova, T.; Kroke, E. *Coord. Chem. Rev.* **2013**, *257*, 2032

(37) Himmelsbach, M.; Vo, T. D. T. *Electrophoresis* **2014**, *35*, 1362

(38) Atalla, V.; Yoon, M.; Caruso, F.; Rinke, P.; Scheffler, M. *Phys. Rev. B* **2013**, *88*, 165122

(39) Blum, V.; Gehrke, R.; Hanke, F.; Havu, P.; Havu, V.; Ren, X.; Reuter, K.; Scheffler, M. *Comput. Phys. Commun.* **2009**, *180*, 2175

(40) Ren, X.; Rinke, P.; Blum, V.; Wieferink, J.; Tkatchenko, A.; Sanfilippo, A.; Reuter, K.; Scheffler, M. *New J. Phys.* **2012**, *14*, 053020

(41) Martin, D. J.; Qiu, K.; Shevlin, S. A.; Handoko, A. D.; Chen, X.; Guo, Z.; Tang, J. *Angew. Chem.* **2014**, *53*, 9240

(42) Berr, M. J.; Wagner, P.; Fischbach, S.; Vaneski, A.; Schneider, J.; Susha, A. S.; Rogach, A. L.; Jäckel, F.; Feldmann, J. *Appl. Phys. Lett.* **2012**, *100*, 223903

(43) Jürgens, B.; Irran, E.; Senker, J.; Kroll, P.; Müller, H.; Schnick, W. *J. Am. Chem. Soc.* **2003**, *125*, 10288

(44) Lotsch, B. V.; Schnick, W. *Chem. Mater.* **2005**, *17*, 3976

(45) Lotsch, B. V.; Schnick, W. *Chem. Mater.* **2006**, *18*, 1891

(46) Horvath-Bordon, E.; Kroke, E.; Svoboda, I.; Fuess, H.; Riedel, R. *New J. Chem.* **2005**, *29*, 693

(47) Komatsu, T. *Macromol. Chem. Phys.* **2001**, *202*, 19

(48) Komatsu, T. *J. Mater. Chem.* **2001**, *11*, 799

(49) Komatsu, T. *J. Mater. Chem.* **2001**, *11*, 802

(50) Irran, E.; Jürgens, B.; Schnick, W. *Solid State Sciences* **2002**, *4*, 1305

(51) Makowski, S. J.; Gunzelmann, D.; Senker, J.; Schnick, W. *Z. Anorg. Allg. Chem.* **2009**, *635*, 2434

(52) Buchanan, G. W.; Crutchley, R. J. *Magn. Reson. Chem.* **1994**, *32*, 552

(53) Lotsch, B. V.; Schnick, W. *Z. Anorg. Allg. Chem.* **2007**, *633*, 1435

(54) Nag, A.; Lotsch, B. V.; Günne, J. S. a. d.; Oeckler, O.; Schmidt, P. J.; Schnick, W. *Chem. Eur. J.* **2007**, *13*, 3512

(55) Fina, F.; Callear, S. K.; Carins, G. M.; Irvine, J. T. S. *Chem. Mater.* **2015**, *27*, 2612

(56) Fina, F.; Ménard, H.; Irvine, J. T. S. *Phys. Chem. Chem. Phys.* **2015**, *17*, 13929

(57) Bera, P.; Priolkar, K. R.; Gayen, A.; Sarode, P. R.; Hegde, M. S.; Emura, S.; Kumashiro, R.; Jayaram, V.; Subbanna, G. N. *Chem. Mater.* **2003**, *15*, 2049

(58) Zhang, F.; Chen, J.; Zhang, X.; Gao, W.; Jin, R.; Guan, N.; Li, Y. *Langmuir* **2004**, *20*, 9329

(59) Nagai, Y.; Hirabayashi, T.; Dohmae, K.; Takagi, N.; Minami, T.; Shinjoh, H.; Matsumoto, S. i. *J. Catal.* **2006**, *242*, 103

(60) Lewera, A.; Timperman, L.; Roguska, A.; Alonso-Vante, N. *J. Phys. Chem. C* **2011**, *115*, 20153

(61) Ohyama, J.; Yamamoto, A.; Teramura, K.; Shishido, T.; Tanaka, T. *ACS Catal.* **2011**, *1*, 187

(62) Akalework, N. G.; Pan, C.-J.; Su, W.-N.; Rick, J.; Tsai, M.-C.; Lee, J.-F.; Lin, J.-M.; Tsaic, L.-D.; Hwang, B.-J. *J. Mater. Chem.* **2012**, *22*, 20977

(63) Parola, V. L.; Kantcheva, M.; Milanova, M.; Venezia, A. M. *J. Catal.* **2013**, *298*, 170

(64) Hwang, C.-P.; Yeh, C.-T. *J. Catal.* **1999**, *182*, 48

(65) Bruijn, F. A. d.; Marin, G. B.; Niemantsverdriet, J. W.; Visscher, W. H. M.; Veen, J. A. R. v. *Surf. Interface Anal.* **1992**, *19*, 537
(66) Kumar, A.; Ramani, V. *ACS Catal.* **2014**, *4*, 1516
(67) Ntais, S.; Isaifan, R. J.; Baranova, E. A. *Mater. Chem. Phys.* **2014**, *148*, 673
(68) Li, L.; Salvador, P. A.; Rohrer, G. S. *Nanoscale* **2014**, *6*, 24
(69) Park, S.; Lee, C. W.; Kang, M.-G.; Kim, S.; Kim, H. J.; Kwon, J. E.; Park, S. Y.; Kang, C.-Y.; Hong, K. S.; Nam, K. T. *Phys. Chem. Chem. Phys.* **2014**, *16*, 10408
(70) Huda, M. N.; Turner, J. A. *J. Appl. Phys.* **2010**, *107*, 123703

**Table of Content Graphic**

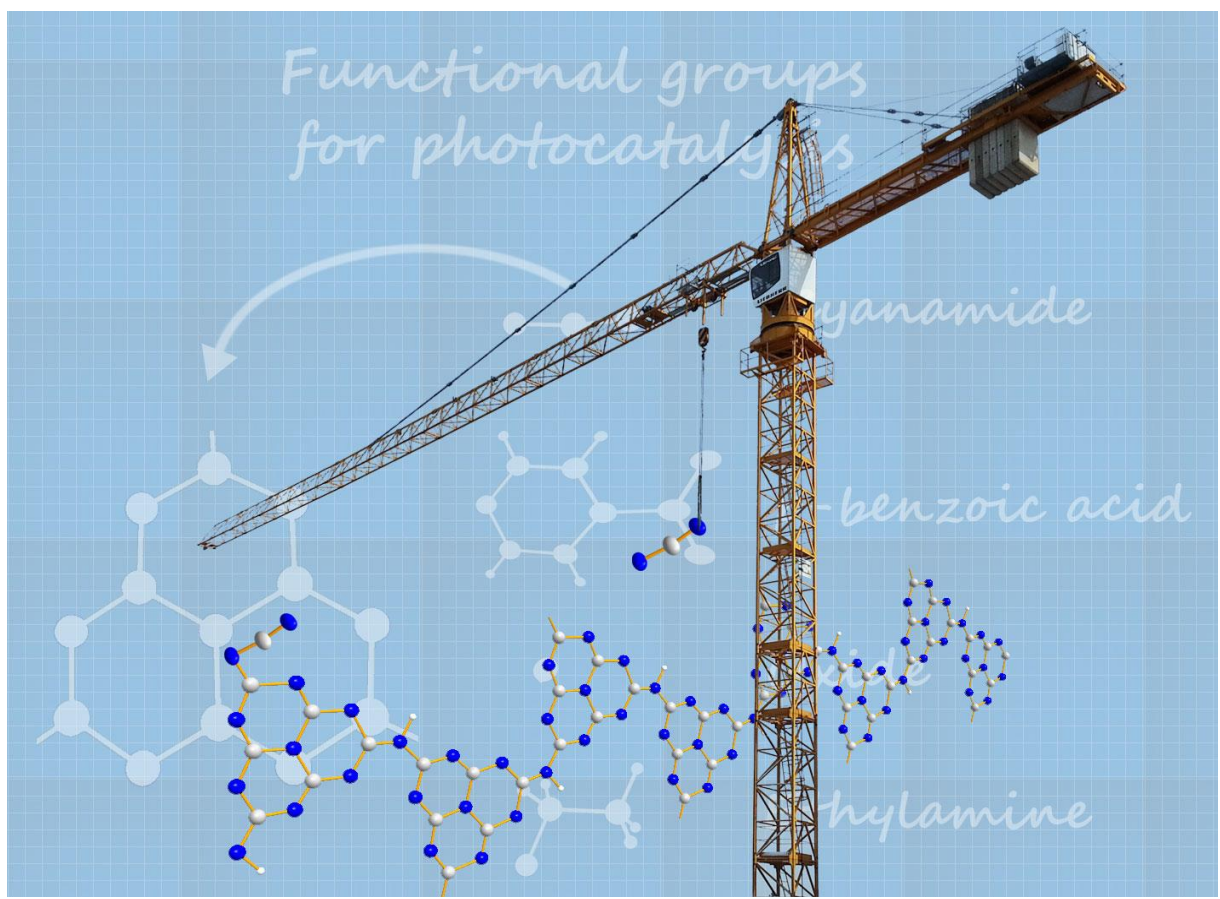